*Article*

# A Sequential Method of Detecting Abrupt Changes in the Correlation Coefficient and Its Application to Bering Sea Climate

**Sergei Rodionov**

Climate Logic LLC, Superior, CO 80027, USA; E-Mail: rodionov.sergei@gmail.com; Tel.: +1-303-900-3115

**Abstract:** A new method of regime shift detection in the correlation coefficient is proposed. The method is designed to find multiple change-points with unknown locations in time series. It signals a possible regime shift in real time and allows for its monitoring. The method is tested on randomly generated time series with predefined change-points. It is applied to examine structural changes in the Bering Sea climate. A major shift is found in 1967, which coincides with a transition from a zonal type of atmospheric circulation to a meridional one. The roles of the Siberian and Alaskan centers of action on winter temperatures in the eastern Bering Sea have been investigated.

## 1. Introduction and Problem Formulation

The correlation coefficient is a principal statistical tool and the most widely used measure of a relationship between two variables [1]. When using this tool, it is assumed that the nature of the relationship is linear and can be modelled by a simple linear regression. Large natural systems, however, such as the climate system, exhibit behaviors that are far more complex and seem to require equally complex, possibly nonlinear, models to simulate them adequately. An alternative view, which is adopted here, is that such systems may be governed by simple rules, but the parameters of those rules are changing, possibly in an abrupt fashion. Indeed, quasi-permanent regimes with abrupt changes between them are characteristic of the climate system and have been observed on different time scales [2–7]. In this paradigm, a major task would be to determine the timing of those abrupt changes, also known as breaks, regime shifts, discontinuities, or change-points. In practice, when new data arrive over time, it is also important to allow for monitoring of those shifts and detect them as soon as possible.

Given this framework, a relationship between two variables can be written in a sequential manner as a dynamic regression model [8]:

$$y_i = x_i \beta_i + e_i, i = 1, \ldots n, n + 1, \ldots, \quad (1)$$

where $x_i$ and $y_i$ are observations at time $i$ of the independent and dependent variables, respectively, $\beta_i$ are the regression coefficients and $e_i$ are the error terms or residuals. The sample $\{(x_1, y_1), \ldots, (x_n, y_n)\}$ is called the historic sample, for which observations are available, and new observations are expected to arrive after time $n$. Let this historic sample be a random sample from a bivariate normal distribution with means $\mu_x, \mu_y$, variances $\sigma_x^2, \sigma_y^2$, and population correlation coefficient $\rho$. The idea is to estimate the regression coefficient just once for the history period $1, \ldots, n$, assuming it is constant during all that time, *i.e.*, $\beta_i \equiv \beta_0$. Alternatively, one can estimate the correlation coefficient instead, because in a simple linear regression, the values of the ordinary least squares (OLS) estimate $\hat{\beta}$ and the Pearson's correlation coefficient $r$ are the same when the time series $x$ and $y$ are normalized by their respective standard deviations $s_x, s_y$. Then a process is devised that captures fluctuations either in estimates or in residuals of a regression model during the monitoring period. Whenever there are excessive fluctuations in these processes, that is, $\hat{\beta}_i$ or $r_i$ for $i > n$ become significantly different from $\beta_0$ or $r_0$, respectively, or the residuals $e_i$ deviate systematically from their zero mean, the null hypothesis of stability is rejected and a regime shift is declared.

This approach works well in statistical quality control applications, when the underlying process is known and the constancy of the correlation coefficient during the history period can be assured [9]. In the case of natural systems, however, there is no guarantee that the data collected came from the same population. On the contrary, it is quite possible that there were already multiple regime shifts during the history period that occurred at unknown times. Therefore, a more appropriate model would be

$$y_i = x_i \beta_j + e_i, i = c_j, c_j + 1, \ldots, c_{j+1} - 1, j = 1, \ldots, m, m + 1, \ldots \quad (2)$$

where each change point $c_j$ is the first point of regime $j$. The initial task here is to find the number of regimes $m$ and locations of change-points $c_j, j = 2, \ldots, m$ before the monitoring phase can begin. To resolve this task, both sequential and non-sequential (also known as retrospective or "ex-post") approaches can be used. The retrospective tests usually fall into one of the two categories: (1) Cumulative sum (CUSUM) and fluctuation tests or (2) $F$-tests, such as the Chow test [8]. The classic CUSUM tests are based on the behavior of recursive residuals $e_i$ [10]. Fluctuation tests are CUSUM-type tests, but deal with changes in estimates of the correlation coefficients $\beta_i$ [8]. While the tests in the first category can be used in both retrospective and monitoring settings, the $F$-tests cannot be applied to monitor out-of-sample stability each time new data arrive [11]. Due to the law of the iterated logarithm, repeated application of such tests yields a procedure that rejects a true null hypothesis of no change with probability approaching one as the number of applications grows [12]. In addition, all retrospective tests have a common problem: the drastic deterioration of the test statistics toward the end of time-series. Therefore, the null hypothesis of constancy of the regression coefficient for the last regime in the historic sample, $i = c_m, \ldots, n$, would be weak and the transition to the monitoring phase would be problematic.

Staying within the sequential framework, one can use subsamples of the data (e.g., in a form of a sliding window as in Leisch *et al.* [13]), assuming that the subsample size is equal or smaller than the length of anticipated regimes. This will improve data homogeneity within the subsamples, but increase the standard error of the correlation coefficient. It is known that the Pearson's $r$ is a biased estimator of ρ, and the bias increases as the sample size decreases [14]. The bias can be substantially reduced by an estimator suggested by Fisher [15] and almost completely eliminated by a related estimator recommended by Olkin and Pratt [16]. A much bigger issue, however, is that even if the correlation coefficient remains constant within a subsample, there might be changes in the means and variances in one or both variables occurring at unknown times. In these circumstances, an accurate estimation of the correlation coefficient directly from data, without addressing this issue first, is not feasible.

A solution suggested here represents a three-step procedure discussed in detail in Section 2. In Section 3, this procedure is tested using synthetic time series with predefined statistics. It is demonstrated that failure to remove regime shifts in the mean and variance first may result in spurious regime shifts in the correlation, whereas true regime shifts may not be detected at all. Section 4 discusses a real world example of structural changes in the Bering Sea climate. This example was previously used by Wang *et al.* [17] who found a shift in the Arctic influence on the region in the 1960s. Since they used smoothed (by 5-yr running means) temperature data series and 25-yr running correlation coefficients as a tool, the timing of the shift was approximate. The shift is reexamined in Section 4 using the proposed method. The results are summarized in Section 5.

## 2. The Three-Step Procedure

The procedure starts with the detection of regime shifts in the means of both $x$ and $y$ time series. It then proceeds with the detection of regime shifts in the variances of the residuals after the first step. Finally, it finds the regime shifts in the correlation coefficient after the residuals are normalized by the standard deviations of each regime found in step two.

### *2.1. Regime Shifts in the Mean*

Most methods related to abrupt structural changes with unknown timing focus solely on such changes in the mean [18]. Rodionov [19] proposed a new method that can detect multiple shifts in the mean level of fluctuations for a given time scale. The method, known as STARS (Sequential *T*-test Analysis of Regime Shifts), has been used in numerous research projects, primarily in climatology (e.g., [20]) and marine ecosystem research [21]. Lately, it has found its way into other areas, such as oceanography [22], hydrology [23], biochemistry [24], metrology [25], and economics [26].

STARS falls into a category of CUSUM-type tests. It uses the $t$-statistic to estimate a threshold, or critical level, for the new regime to be detected. In this forward-looking approach, cumulative sums of deviations are calculated not from the mean of the "current" regime, but from that critical level for a new regime. This makes it easier to automate the decision process when testing a potential change-point. The critical level for a new regime $\bar{x}_{cr}$ is defined as $\bar{x}_{cr}^{\uparrow} = \bar{x}_j + \delta$ (if the shift is up), or $\bar{x}_{cr}^{\downarrow} = \bar{x}_j - \delta$ (if the shift is down), where $\bar{x}_j$ is the mean of the "current" regime and

$$\delta = t\sqrt{2\bar{s}_l^2/l} \tag{3}$$

Here, $t$ is the value of Student's $t$-distribution with $2l - 2$ degrees of freedom for a given probability level $p$, $l$ is the cut-off length of the regimes (somewhat similar to the cut-off point in low-pass filtering), and $\bar{s}_l^2$ is the average variance for running $l$-point intervals in the time series. If the value of a newly arrived observation at time $i$ is outside the interval $[\bar{x}_{cr}^{\downarrow}, \bar{x}_{cr}^{\uparrow}]$, it is marked as a potential change-point $c_p$, which triggers the calculation of the regime shift index (RSI) for this point

$$\text{RSI} = \frac{1}{l\bar{s}_l} \sum_{i=c_p}^{k} (x_i - \bar{x}_{cr}), k = c_p, c_p + 1, \ldots, c_p + l - 1 \qquad (4)$$

Equation (4) shows that the next $l - 1$ points after $c_p$ are used to test the null hypothesis of no regime shift at $c_p$. If the RSI retains its sign during this test, then the null hypothesis is rejected, and $c_p$ becomes a true change point $c_{j+1}$. If at any time the RSI changes its sign, the test for $c_p$ stops. The value of $x_i$ at $i = c_p$ is now considered just a random fluctuation that belongs to the "current" regime $j$. The mean value for the regime, $\bar{x}_j$, is recalculated, and the search for the next change-point continues.

In the case of autocorrelation, or "red noise", in the time series, it is important to filter it out, using a procedure known as "prewhitening". When a time series contains both the autocorrelation and regime shifts, it is impossible to estimate the autocorrelation coefficient using the entire historic sample. It can be done using subsamples of the size smaller than $l$. It is known, however, that the smaller the sample, the larger the bias of the OLS estimator of the autocorrelation coefficient [27]. The bias can be substantially reduced using one of the two techniques: (1) MPK, named after Marriott and Pope [28] and Kendall [29]; or (2) IP4 (Inverse Proportionality with 4 corrections), as described in [30]. The IP4 technique is based on the assumption that the bias is approximately inversely proportional to the subsample size and is always negative [31]. Since the residual bias is also inversely proportional to the subsample size, three additional corrections practically eliminate the bias. Monte Carlo simulations showed that for subsample sizes smaller than 10, IP4 substantially outperforms MPK in terms of both the magnitude of the bias and variability of the estimates [30]. Both these techniques became part of the STARS package.

After all the regime shifts in the mean are detected in both $x$ and $y$, the stepwise trends are removed. The residual time series $x'$ and $y'$, which represent deviations from the respective stepwise trends, are then passed to step two for detection of regime shifts in the variance.

## 2.2. Regime Shifts in the Variance

The procedure for detecting regime shifts in the variance is similar to the one for the mean, except that, instead of the $t$-statistic, it uses the $F$-statistic to estimate the critical variance for the new regime. The $F$-test consists of comparing the ratio of sample variances for two adjacent regimes with the critical value $F_{cr}$:

$$F = \frac{s_j^2}{s_{j+1}^2} \lessgtr F_{cr}. \qquad (5)$$

Here $F_{cr}$ is the value of the $F$-distribution with $v_1$ and $v_2$ degrees of freedom (where $v_1 = v_2 = l - 1$) and a significance level $p$ (two-tailed test), $F_{cr} = F(p/2, v_1, v_2)$. For the new regime to be statistically different from the "current" regime, variance $s_{j+1}^2$ should be greater than critical variance $s_{cr}^{2\uparrow}$, if the variance increases, or smaller than $s_{cr}^{2\downarrow}$, if the variance decreases, where

$$s_{cr}^2 = \begin{cases} s_{cr}^{2\uparrow} = s_j^2 F_{cr}, \\ s_{cr}^{2\downarrow} = s_j^2 / F_{cr} \end{cases} \quad (6)$$

If at the "current" time $i$, the value of $x_i'^2$ is greater than $s_{cr}^{2\uparrow}$ or smaller than $s_{cr}^{2\downarrow}$, this point is marked as a potential change-point $c_p$, and the subsequent $l - 1$ data points are used to test the null hypothesis of no regime shift. The decision rule is similar to the one used for shifts in the mean, but based on the residual sum of squares index (RSSI):

$$RSSI = \frac{1}{l} \sum_{i=c_p}^{k} (x_i'^2 - s_{cr}^2), k = c_p, c_p + 1, \ldots, c_p + l - 1 \quad (7)$$

If at any point during this testing period the index turns negative in the case of increasing variance, or positive in the case of decreasing variance, the test for $c_p$ stops, because the null hypothesis cannot be rejected. The data point $x_i'^2$ is included in the "current" regime $j$, and its variance $s_j^2$ is recalculated. Otherwise, the null hypothesis is rejected, and $c_p$ becomes a true change-point $c_{j+1}$.

After finding all the regime shits in both $x'$ and $y'$, the time series are normalized by dividing each value by the standard deviation for the corresponding regime. The normalized time series $x^*$ and $y^*$ are then passed to step three for detection of regime shifts in the correlation.

*2.3. Regime Shifts in the Correlation Coefficient*

The detection of regime shifts in the correlation coefficient between $x^*$ and $y^*$ is based on the formula for the variance of their sum [1]:

$$s_{x^*+y^*}^2 = s_{x^*}^2 + s_{y^*}^2 + 2rs_{x^*}s_{y^*} \quad (8)$$

Since in our case the variables are normalized, *i.e.*, $\bar{x}^* = \bar{y}^* = 0$ and $s_{x^*}^2 = s_{y^*}^2 = 1$, the formula becomes

$$s_{x^*+y^*}^2 = 2(1 + r) \quad (9)$$

Equation (9) shows that, as the correlation coefficient increases, the variance also increases from a minimum of 0 at $r = -1$ to a maximum of 4 at $r = 1$. To detect abrupt shifts in $r$, the technique described in step 2, can be applied to a series of the sums $x^* + y^*$. It can also be applied to a series of the differences $x^* - y^*$ since

$$s_{x^*-y^*}^2 = 2(1 - r) \quad (10)$$

For practical reasons, the detection test is performed for both $x^* + y^*$ and $x^* - y^*$ series. Our experiments show that when the difference between the correlation coefficients for two adjacent regimes is statistically significant at $p \leq 0.05$, the detected change-points most often are the same for both series. For $p > 0.05$, the change points may differ by a few time steps or may be detected only in one of the series. In the case of two competing potential change-points, the decision about which one to select is based on their $p$-values computed using the Fisher r-to-z-transformation [32]. The change point with the lower $p$-value is chosen as the final one.

## 3. An Example with Synthetic Time Series

The above three-step algorithm was implemented as a software package called the Sequential Regime Shift Detector (SRSD). It is written in VBA for MS Excel and is available at [33].

To illustrate the software capabilities, two correlated random time series of size $n = 70$ were generated from the bivariate normal distribution with $\mu_x = \mu_y = 0$ and $\sigma_x^2 = \sigma_y^2 = 1$. The change-point for the correlation coefficient from regime 1 to regime 2 was set in the middle of the series at $c_2 = 36$, with $\rho_1 = -0.6$ in the first half and $\rho_2 = 0.6$ in the second half. Then the shifts in the variance were added to time series $x$ from $\sigma_1^2 = 1$ to $\sigma_2^2 = 9$ at $c_2 = 51$ and to time series $y$ from $\sigma_1^2 = 9$ to $\sigma_2^2 = 1$ at $c_2 = 21$. Finally, the shifts in the mean were superimposed to $x$ from $\mu_1 = -1$ to $\mu_2 = 1$ at $c_2 = 26$ and to $y$ from $\mu_1 = 1$ to $\mu_2 = -1$ at $c_2 = 41$. The resultant time series are presented in Figure 1.

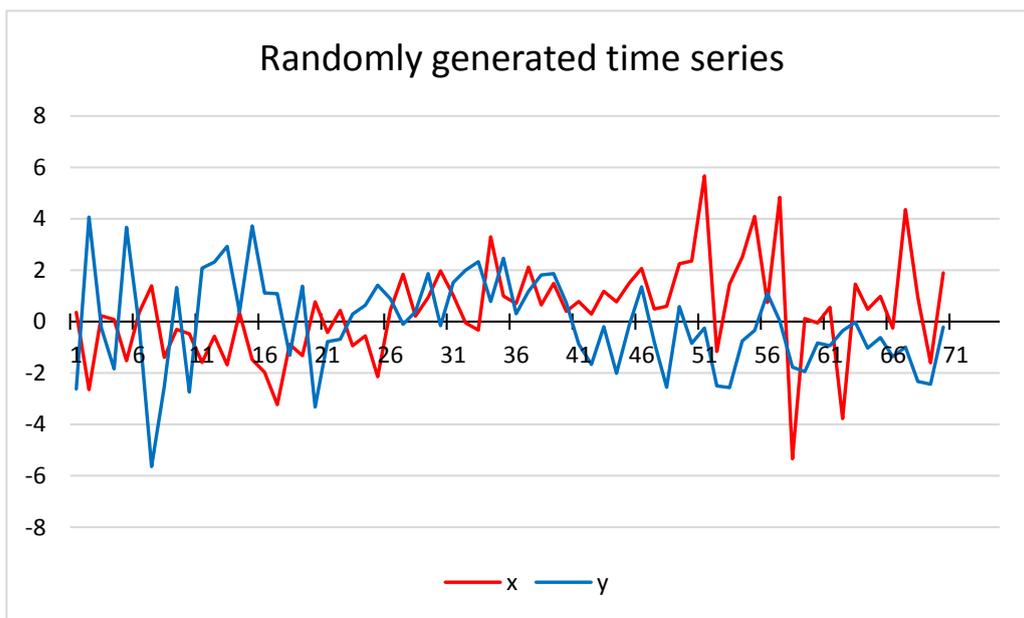

**Figure 1.** Randomly generated time series with shifts in the correlation coefficient, variance and mean as described in the text.

The addition of regime shifts in the variance and, particularly, in the mean significantly changes the direct OLS estimates of ρ. As shown in Figure 2, in the presence of those shifts, sample correlation coefficient $r$ tends to underestimate ρ by as much as 0.6. In some other cases, when the shifts in the mean and variance are placed differently, it may lead to an overestimation of ρ. Therefore, in order to detect shifts in the correlation coefficient accurately, it is necessary to remove shifts in the mean and variance first.

The SRSD starts by detecting shifts in the mean. In this example, the target significance level and the cut-off length are set at $p = 0.05$ and $l = 20$. It is always good to know an approximate length of the regimes in time series to set these two parameters right, but since there is only one strong regime shift in the mean in each of the series, some variations in $p$ and $l$ do not change the results. Figure 3a,b show that the detected change points $\hat{c}_2 = 26$ and $\hat{c}_2 = 41$ in $x$ and $y$, respectively, coincide with the theoretical change points. The calculated $p$–values in both cases are much lower than 0.05.

After the stepwise trends are removed, the detection of regime shifts in the variance is performed using the same target significance level $p = 0.05$ and the cut-off length $l = 20$. The increase in the variance in $x$ (Figure 3c) and its decrease in $y$ (Figure 3d) were both accurately detected at $\hat{c}_2 = 51$ and $\hat{c}_2 = 21$, respectively.

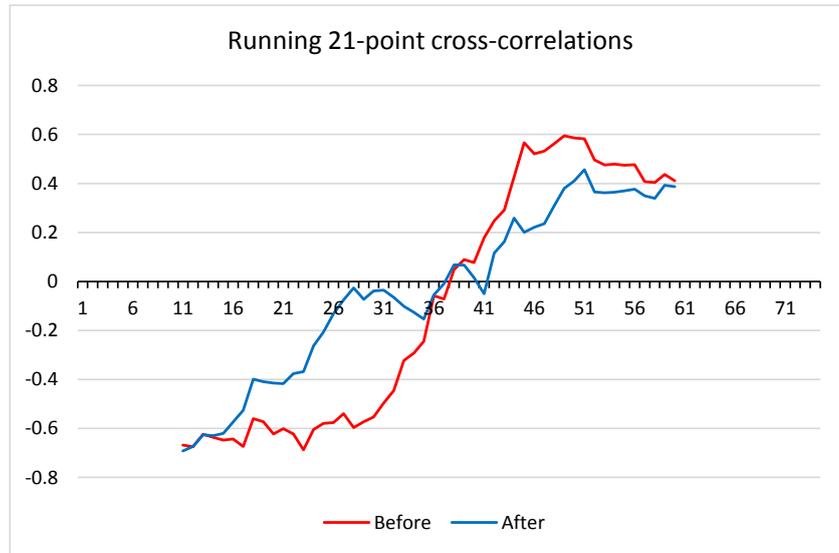

**Figure 2.** Running 21-point cross-correlations before and after the shifts in the variance and mean are added to the time series.

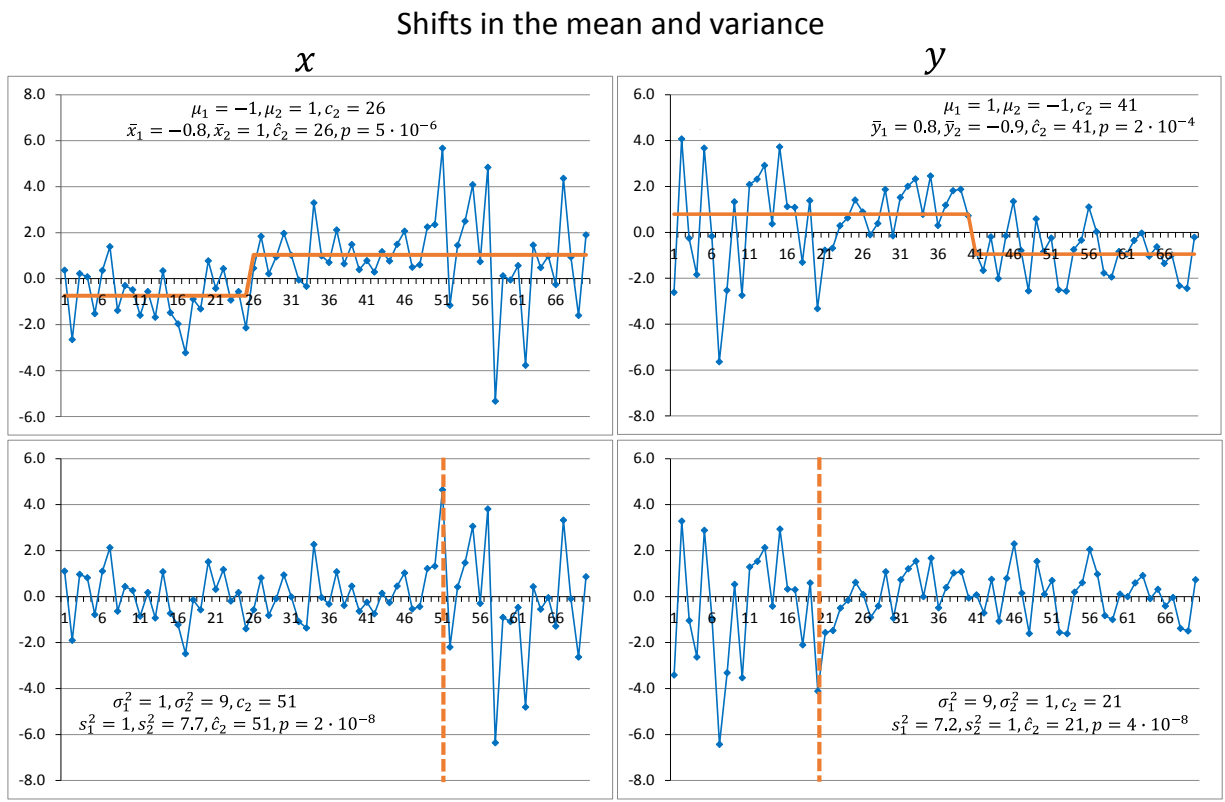

**Figure 3.** Regime shifts detected in the mean (**top row**) and variance (**bottom row**) in time series $x$ (**left column**) and $y$ (**right column**). In all four cases the detection is performed with the target significance level $p = 0.05$ and cut-off length $l = 20$.

In its final step, the SRSD uses the normalized time series to detect regime shifts in the correlation coefficient. As shown in Figure 4a, the regime shift in our synthetic time series is found at $\hat{c}_2 = 36$, which is precisely where it is supposed to be. Interestingly, when the first two steps are skipped, that is the regime shifts in the mean and variance are not removed, a shift in $r$ is incorrectly detected at $\hat{c}_2 = 21$ (Figure 4b). Although the difference in $r$ before and after this spurious shift appears to be statistically significant (the exact $p$-value cannot be calculated due to violation of the assumptions for $r$), its timing is way off.

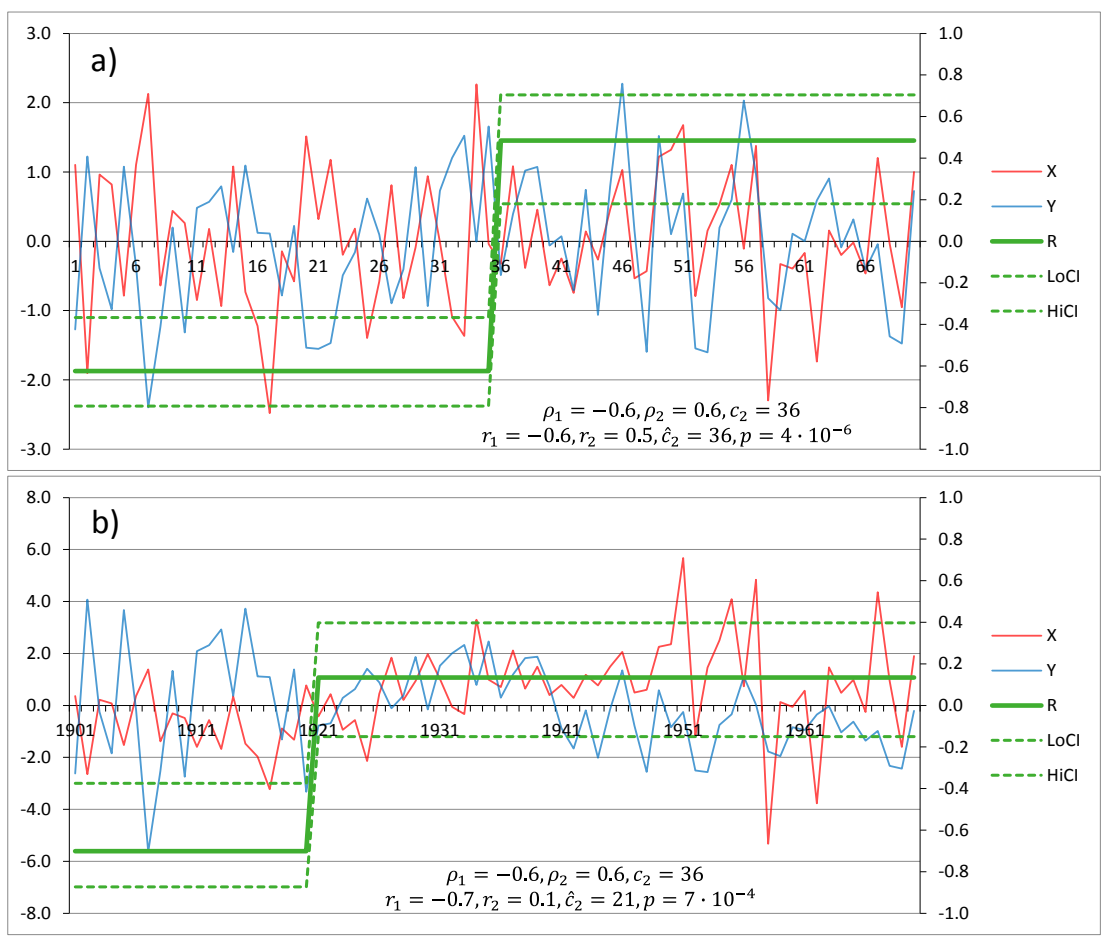

**Figure 4.** Regime shifts in the correlation coefficient (**a**) correctly detected at $i = 36$, after shifts in the mean and variance are removed; and (**b**) a spurious shift at $i = 21$, when the SRSD is applied directly to the original series in Figure 1.

## 4. Change in the Climatological Structure of the Bering Sea

Wang *et al.* [17] reported a change of unknown origin in the climatological structure of the Bering Sea that occurred sometime in the 1960s. Using the running 25-yr correlation coefficients between winter (DJFM) surface air temperatures (SAT) at two Alaskan stations, Barrow (71.3 °N, 156.8 °W) and St. Paul (57.1 °N, 170.2 °W), they found a shift from about 0.2 to 0.7 between the 1940s and 1980s. They concluded that the shift was statistically significant at the 95% confidence level, based on their Monte Carlo simulations.

With the approach they used, it was difficult to determine the exact timing of the shift, especially when they smoothed the SAT time series with running 5-yr means, which only increased the autocorrelation in the series. In addition, a visual inspection of their Figure 2a reveals a strong shift in SAT at Barrow, from a cold regime before the late 1970s to a warm regime thereafter. As shown in Section 3, it may lead to large errors in estimation of the correlation coefficient.

The station SAT data used in this study was obtained from GISS [34]. To facilitate a comparison between SATs at Barrow and St. Paul, the data for both stations is presented as anomalies from the 1971–2000 base period normalized by the standard deviations for the same period. Figure 5a shows that SAT at Barrow experienced an abrupt, and statistically highly significant, shift upward in 1978, the $p$-value of which was $8 \cdot 10^{-7}$. It coincided with a major step in the North Pacific climate in the late 1970s [35]. Interestingly, no shift at that time was detected in SAT at St. Paul, despite its proximity to the North Pacific Ocean (Figure 5b). It should be noted that, using the same method, a weak shift in 1977 was detected for that station in [36]. The discrepancy can be explained by the difference in the data sources used, that is, the raw data from NCDC in [36] and adjusted data from GISS here. Apparently, the homogeneity adjustment technique used in GISS made the 1977 shift in SAT at St. Paul even weaker. In any case, the absence of a pronounced shift in SAT at St. Paul is a reflection of the complex climate dynamics in the region and the lack of a clear-cut connection with the major modes of climate variability in the Pacific. As shown in [37], the correlation coefficients between winter SAT at St. Paul and such indices as the Pacific Decadal Oscillation, Aleutian Low Pressure, Pacific-North American, and Southern Oscillation are all statistically insignificant.

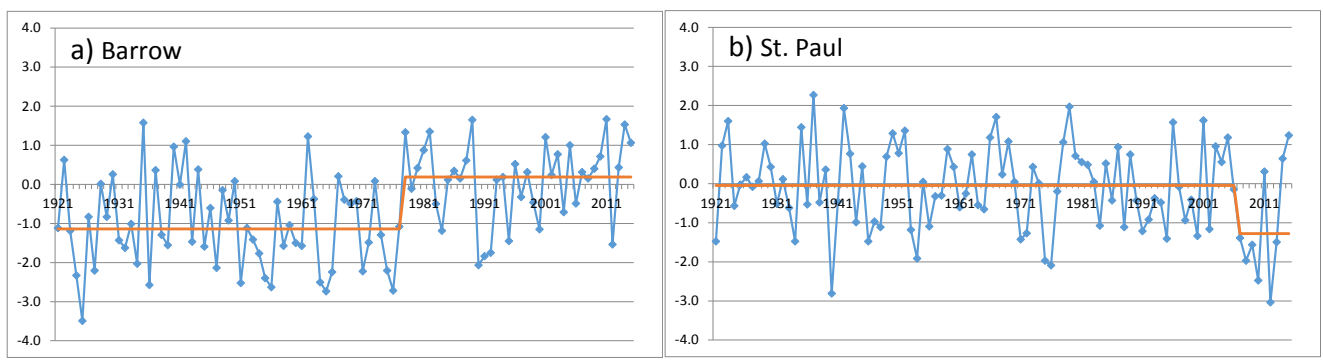

**Figure 5.** Regime shifts in the mean winter (DJFM) surface air temperatures (SATs) at (**a**) Barrow and (**b**) St. Paul, 1921–2015. The detection was performed using the target significance level $p = 0.1$ and cut-off length $l = 15$. The autocorrelation coefficients, estimated using the IP4 method [30], are close to zero in both series. The $p$-values for the change-points in 1978 for Barrow and in 2007 for St. Paul are $8 \cdot 10^{-7}$ and 0.07, respectively.

The relative stability of temperature fluctuations at St. Paul was interrupted in 2007, with the onset of a very cold period that lasted through 2013 and included a record cold winter of 2012. This period was characterized by a deep and persistent low-pressure center located in the Gulf of Alaska [38], which implies the advection of cold continental air over the southeastern Bering Sea [36].

When the regime shifts in the mean were removed, the maximum $r$ between SAT residuals at Barrow and St. Paul for any 25-yr sample decreased from 0.7 to 0.6. Since no regime shifts in the variance were detected, step two in the three-step procedure was skipped, and the test for shifts in the correlation coefficient was performed using the residuals after the first step. The target significance level and cut-off length were the same as for the tests of shifts in the mean (*i.e.*, target $p = 0.1$ and $l = 15$).

Figure 6 shows two strong correlation regimes, 1921–1939 ($r = 0.59$, 90% confidence interval: 0.25–0.79) and 1967–2015 (0.69, 0.54–0.80), separated by a period with no correlation between Barrow and St. Paul. The $p$-values for the regime shifts in 1940 and 1967 are 0.04 and 0.001, respectively. Here the focus will be on the latter shift due to its higher statistical significance and more readily available meteorological data to explore it further.

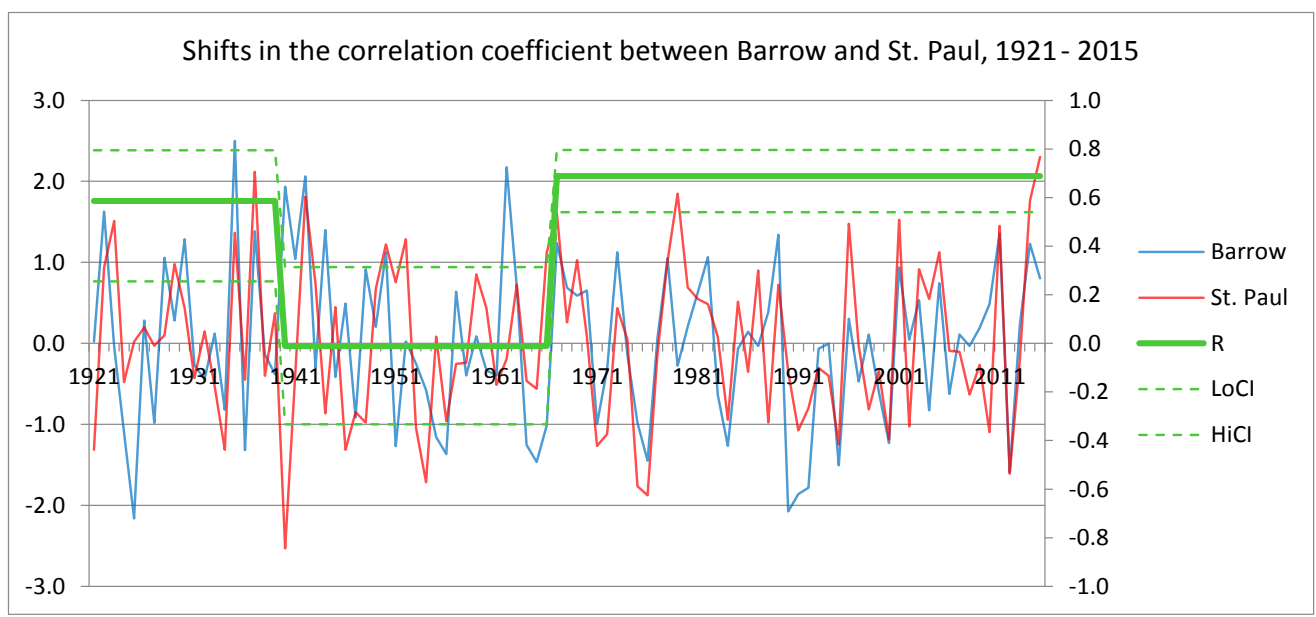

**Figure 6.** Regime shifts in the correlation coefficient between winter SATs at Barrow and St. Paul detected using the target significance level $p = 0.1$ and cut-off length $l = 15$. The $p$-values for the change-points in 1940 and 1967 are 0.04 and 0.001, respectively. The correlation coefficients, along with their 90% confidence intervals, for the three detected regimes are $0.59\ (0.25 - 0.79)$ for 1921–1939, $-0.01\ (-0.33 - 0.31)$ for 1940–1966, and $0.69\ (0.54 - 0.80)$ for 1967–2015.

Baines and Folland [5] reported a number of rapid climate changes in various parts of the globe centered on the late 1960s. In the Bering Sea region, that was the time when atmospheric circulation changed from a generally zonal flow to a meridional pattern [17]. Wang *et al.* [17] demonstrated that Arctic and Pacific air had fewer meridional excursions before the late 1960s. They alluded, however, that since the increase in the north/south correlation structure did not begin near the well-known shift in the late 1970s, but about a decade earlier, it might be more related to changes in the Siberian High.

The Siberian High is part of the so-called Siberian-Alaskan Index (SAI) designed specifically to characterize the effect of atmospheric circulation on thermal conditions in the Bering Sea. The SAI is defined as a difference between the mean winter (DJFM) normalized 700-hPa anomalies in two regions, Siberia (55 N–70 N, 90 E–150 E) and Alaska/Yukon (60 N–70 N, 130 W–160 W). Positive (negative) values of the index indicate anomalously strong north-westerly (south-easterly) winds and colder (warmer) than normal winters in the Bering Sea. The SAI is available at the Bering Climate web site [39].

It should be emphasized that the SAI is not a dipole, as, for example, the North Atlantic Oscillation. The two parts of the SAI, the Siberian Index (SI) and Alaskan Index (AI), are completely uncorrelated, and therefore, represent two independent sources of influence on Bering Sea climate. The SI reflects the strength of the Siberian High (Figure 7a) and the advection of cold Siberian air into the Bering Sea. Judging from the correlations between the SI and winter temperatures computed for the entire period of observations, 1949–2015 (Figure 7c), its effect on winter conditions in the Bering Sea is rather modest. Note that the correlation maps in Figure 7 are computed for the first differences rather than for absolute values. Taking first differences serves as a high-pass filter that practically eliminates the effect of regime shifts in the mean on the correlation coefficient.

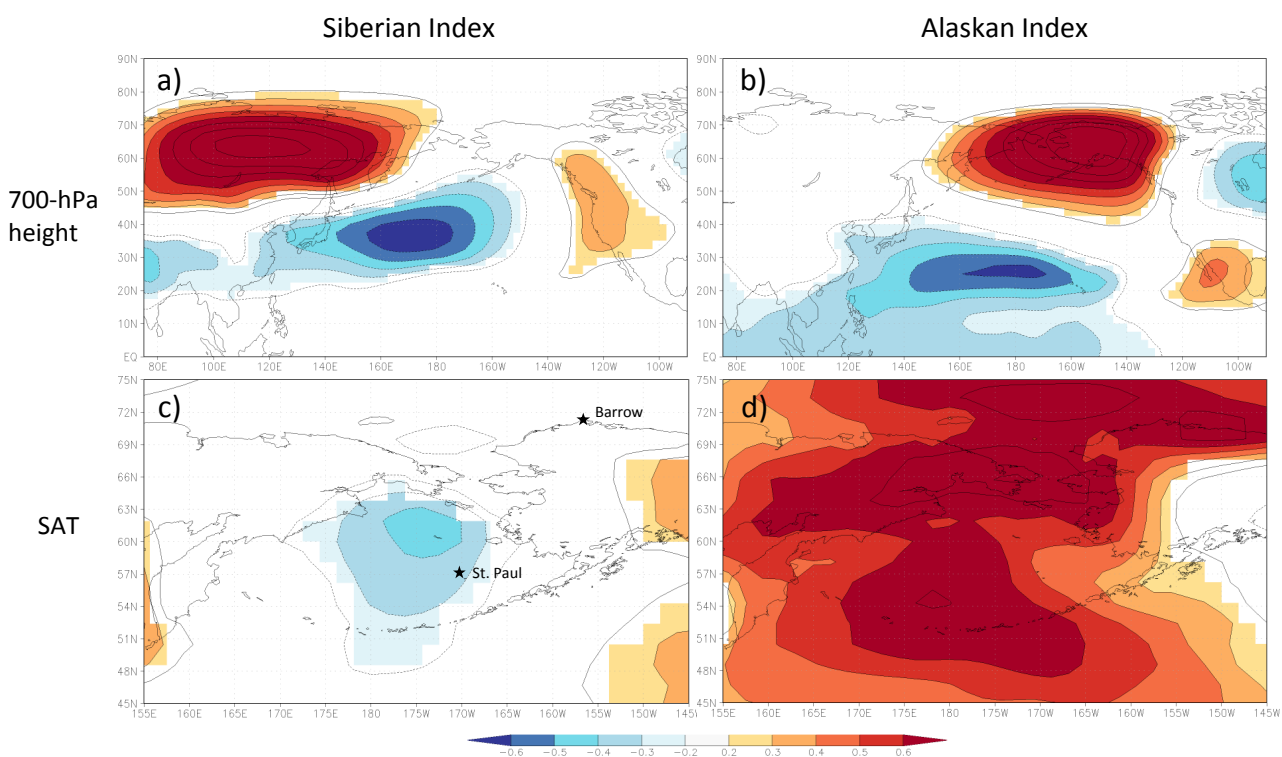

**Figure 7.** Correlation maps for the Siberian (**left column**) and Alaskan (**right column**) indices in the mean winter (DJFM) 700-hPa (**a,b**) height and (**c,d**) SAT fields. The correlation coefficients were calculated using the entire dataset, 1949–2015. The first differences filter ($\Delta x_i = x_i - x_{i-1}$) was applied. Areas where the correlation coefficients are statistically significant at the 90% confidence level are colored.

The 700-hPa correlation map for the AI is shown in Figure 7b. Fang and Wallace [40] demonstrated the importance of a high-pressure center over Alaska for thermal conditions in the Bering Sea. Using the singular value decomposition (SVD) technique for sea ice concentration in the North Pacific and the hemispheric 500-hPa field, they showed that the leading SVD mode was characterized by a dominant center of action over Alaska. Blocking over Alaska (positive AI values) prevents storms from entering the Gulf of Alaska and redirects them into the Bering Sea. These storms bring warm Pacific air and push the ice edge northward. Negative AI values are indicative of advection of cold Arctic air and rapid advance of ice edge southward. A comparison of the SAT correlation maps (Figure 7c,d) shows that the linear effect of the Alaskan center of action on SATs in the Bering Sea is much stronger than that of the Siberian High.

The AI is strongly correlated with SAT at Barrow as well, which can be seen in Figure 7c. For the entire period of observations, 1949–2015, the correlation coefficient between the AI and residuals of SAT at Barrow, after removing the stepwise trend, is 0.61.

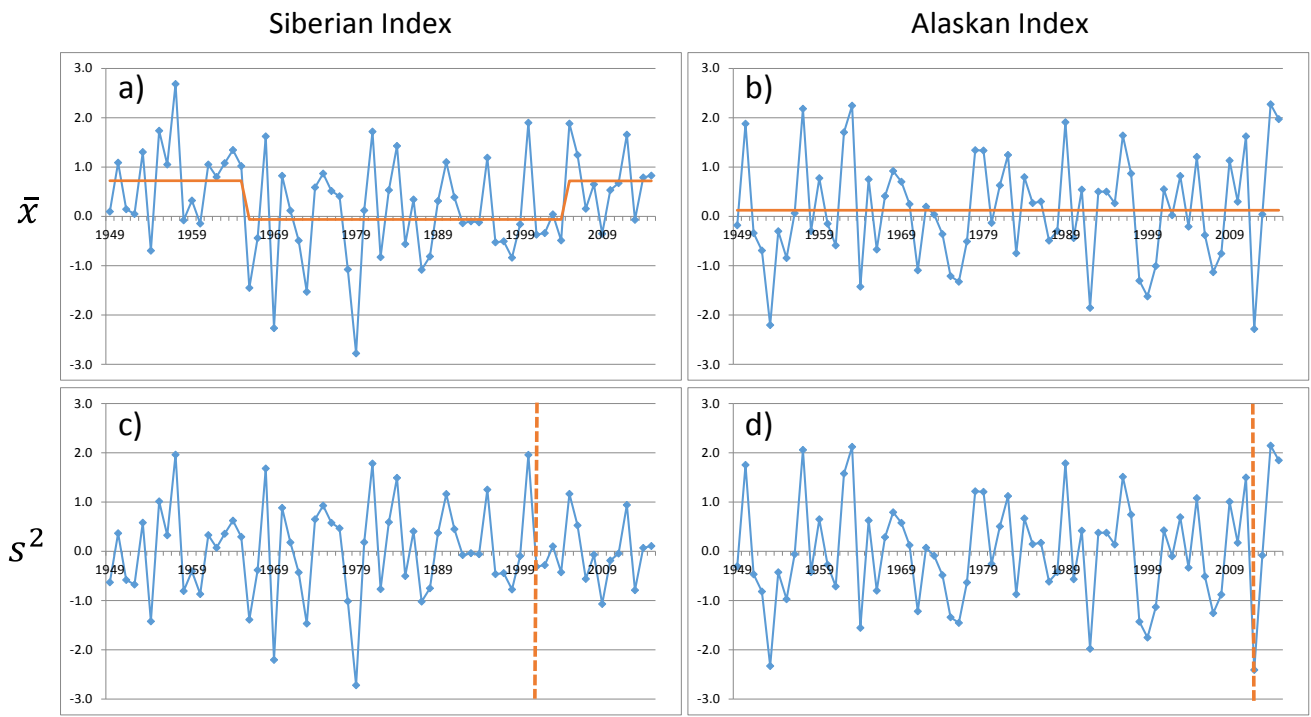

**Figure 8.** Regime shifts in the mean (top row) and variance (bottom row) for the Siberian (left column) and Alaskan (right column) indices. In all four cases, the target significance level and cut-off length were set at 0.1 and 15, respectively. The change-points and their *p*-values (in parentheses) are: (**a**) 1966 (0.003), 2005 (0.006); (**b**) none detected; (**c**) 2001 (0.04); and (**d**) 2012 (not calculated due to a small number of points after the shift).

The results of regime shift detection in the mean and variance in the SI and AI are presented in Figure 8. Prior to 1966, the SI was mostly positive, indicating a strong Siberian High (Figure 8a). Another period of predominantly positive SI started in 2005. No shifts in the mean were detected for the AI (Figure 8b). As for the variance, a statistically significant decrease in the SI fluctuations since 2001 was detected (Figure 8c). In contrast, the AI index experienced a remarkable increase in its fluctuations in recent years, when it jumped from a record low value in 2012 to a record high value in 2014 (Figure 8d). However, due to a small number of observations in this high variance regime, normalization for the AI was not performed.

The results of regime shift detection in the correlation coefficient between SAT at St. Paul and the two atmospheric circulation indices, SI and AI, are presented in Figure 9. Although the effect of the SI on temperature in the Bering Sea appeared to be rather weak overall (Figure 7c), there was a strong correlation regime from 1966 through 1997, when $r$ reached −0.54, a statistically significant value at the 99.8% confidence level (Figure 9a). The linear relationship between the AI and SAT at St. Paul (Figure 9b) was weak at the beginning of the series (1949–1967), but then $r$ jumped from 0.27 to 0.70. It means that about 50% of SAT variance at St. Paul during the 1968–2015 period can be explained by the AI alone.

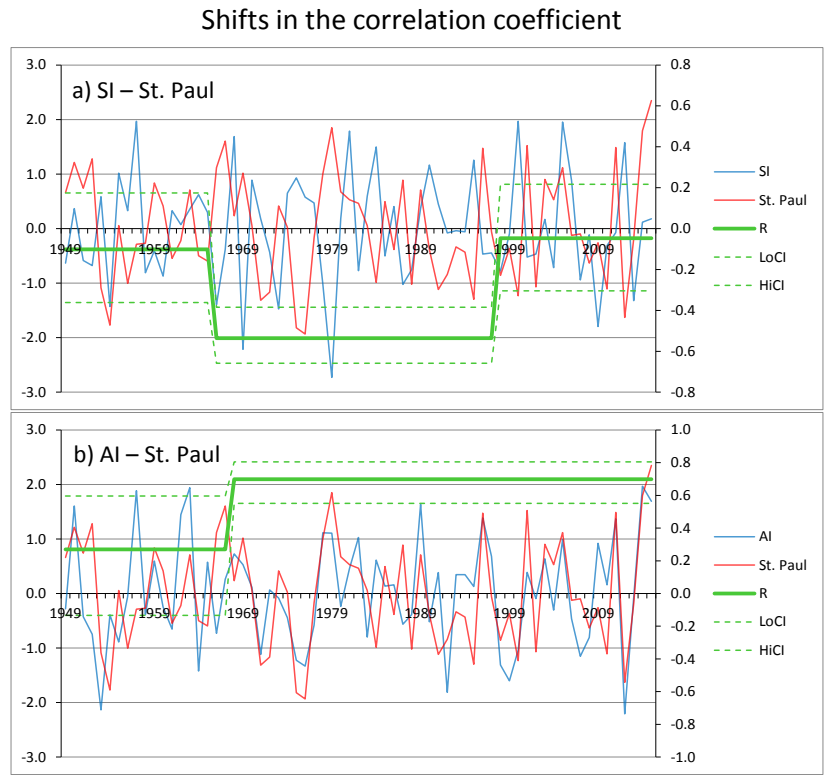

**Figure 9.** Regime shifts in the correlation coefficient between SAT at St. Paul and (**a**) Siberian and (**b**) Alaskan indices. The target significance level and cut-off length were set at 0.3 and 15, respectively, for the pair SI–St. Paul, and 0.1 and 15 for the pair AI–St. Paul. The change-points and their $p$-values (in parentheses) are: (a) 1966 (0.13); 1998 (0.09) and (b) 1968 (0.05).

It is interesting to compare atmospheric circulation patterns for the correlation regimes in Figure 9. As shown in Figure 10, the most prominent feature of atmospheric circulation during the strong correlation regime for the SI, 1966–1997, is a strong Siberian—North Pacific (NP) dipole in the 700-hPa field, with the correlation coefficient between the two centers reaching 0.8 (Figure 10c). During a positive phase of the dipole (SI+, NP−), the pressure gradient between the Siberian High and Aleutian low increases that leads to an enhanced advection of Arctic air directed into the Bering Sea. During its negative phase (SI−, NP+), the advection of Arctic air decreases, while the advection of warm Pacific air increases, apparently due to a more active Siberian storm track [36]. The NP center is weaker during the weak correlation regime of 1949–1965 (Figure 10a) and completely disappears during the recent weak correlation regime of 1998–2015 (Figure 10e).

Figure 11 shed some light on why the correlation coefficient between the AI and SAT at St. Paul was low during the period of 1949–1967. A comparison of Figure 11a,c shows that during the earlier period, the Alaskan center of action was expanded westward. In that situation the advection of warm Pacific air (AI+), or cold Arctic air (AI−) was directed into the western part of the Bering Sea, where the correlation coefficients with SATs reached 0.8, whereas they were statistically insignificant in the eastern part (Figure 11b). Note that the correlation between the AI and SAT at Barrow remained strong during this earlier regime, and that the shift in the correlation for the pairs AI–St. Paul and Barrow—St. Paul occurred almost at the same time. This indicates that the primary reason for the break in the correlation for the latter pair in 1967 is likely to be the westward expansion and subsequent contraction of the Alaskan center of action, and not the change in the Arctic influence as suggested by Wang *et al.* [17].

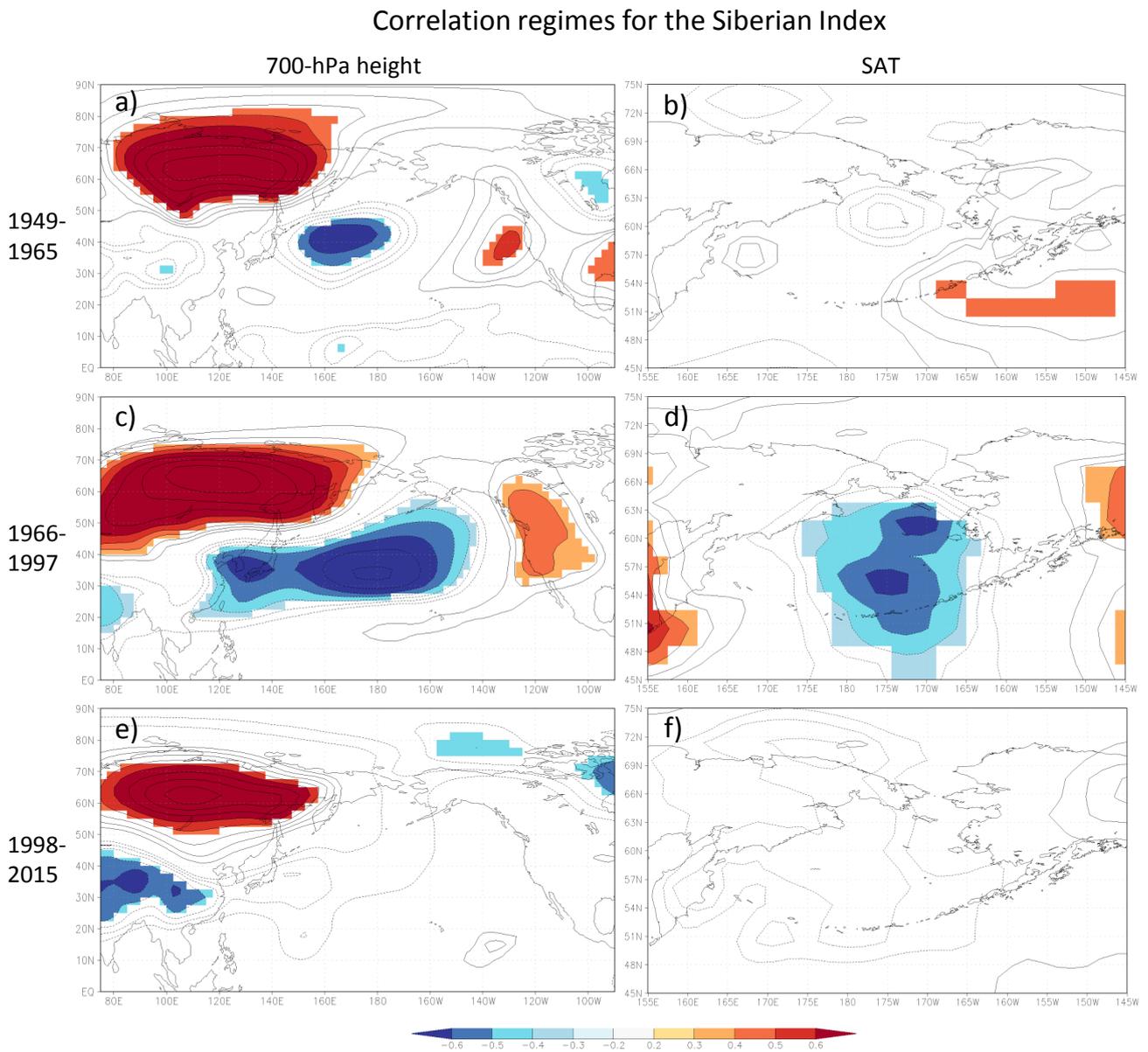

**Figure 10.** Correlation maps for the Siberian index in the mean winter 700-hPa height (**left column**) and SAT (**right column**) fields for three correlation regimes: (1) (**a,b**) 1949–1965; (2) (**c,d**) 1966–1997; and (3) (**e,f**) 1998–2015. The first differences filter was applied. Areas where the correlation coefficients are statistically significant at the 90% confidence level are colored.

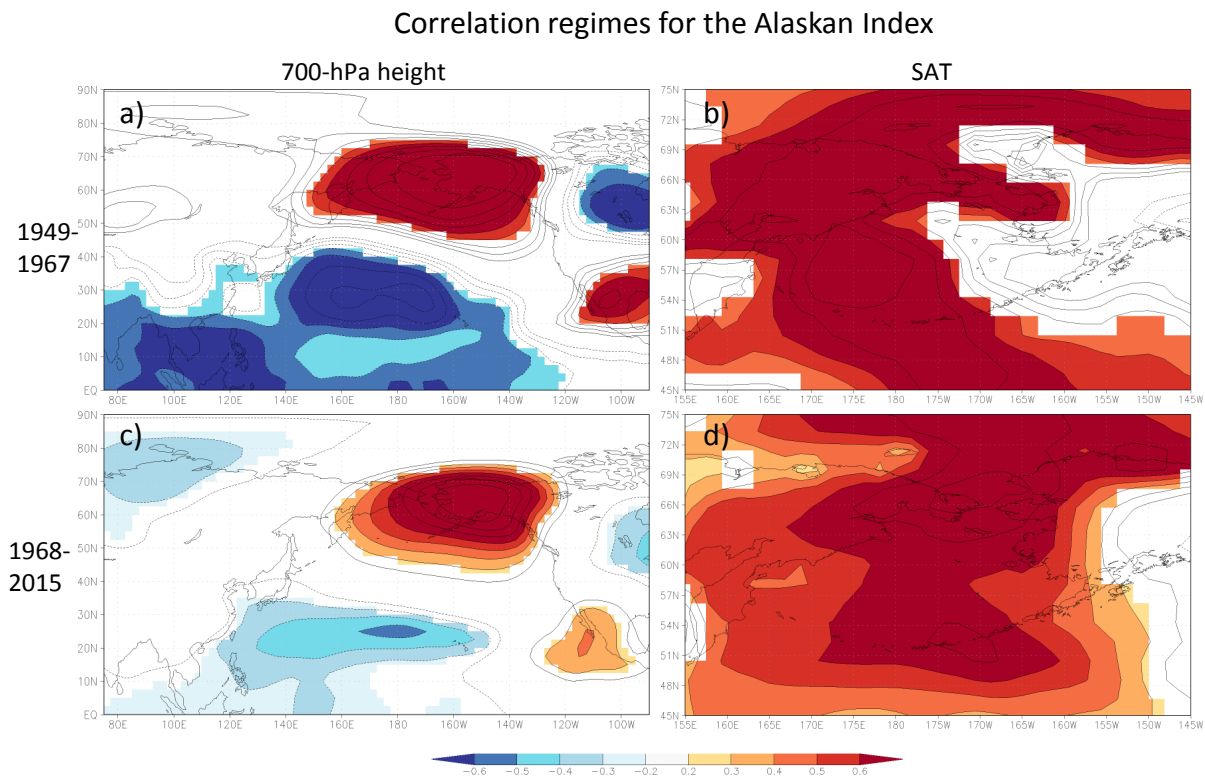

**Figure 11.** Same as Figure 10, except for Alaskan index and two correlation regimes: (1) (**a**,**b**) 1949–1967 and (2) (**c**,**d**) 1968–2015.

## 5. Conclusions

A new method of regime shift detection in the correlation coefficient is proposed. It is capable of automatic detection of multiple change-points at unknown time. It also allows early warning and monitoring of regime shifts. The method is built on previous works by the author [19,30,41], which describe algorithms of sequential regime shift detection in the mean and variance, as well as the prewhitening procedure to eliminate the effect of red noise. Using the synthetic time series, this paper demonstrates that in the presence of shifts in the mean and variance, the direct OLS estimation of the correlation coefficient becomes unreliable. Therefore, a three-step procedure is suggested, which detects and removes the regime shifts in the mean and variance first, and then uses the sequential $F$-test applied to the sums $x^* + y^*$ and differences $x^* - y^*$ of the residuals to detect regime shifts in the correlation coefficient.

The SRSD software based on this procedure has been applied to detect structural changes in the Bering Sea climate. It is shown that a major shift in the correlation between winter SATs at Barrow and St Paul occurred in 1967, which coincides with the change in the atmospheric circulation from a zonal to meridional pattern, as described in [17].

The role of the Siberian and Alaskan centers of action, which represent two independent sources of influence on winter thermal conditions in the Bering Sea, has been investigated. Although the overall effect of the Siberian center is much weaker than the Alaskan one, there was a period (1966–1997) when the correlation coefficient between the SI and SAT at St. Paul reached −0.59, which was statistically significant at the 99.98% confidence level. The principal feature during that strong correlation regime was a well-expressed North Pacific center of action, which acted in a coordinated fashion with the Siberian center.

The effect of the Alaskan center of action on the eastern Bering Sea was relatively weak during the earlier period, 1949–1967. This can be explained by a westward expansion of the center during that period. As a result, fluctuations in the strength of the AI had more effect on the western Bering Sea. After the regime shift in 1968, almost 50% of the variance of winter SAT at St. Paul can be explained by the AI alone. This westward expansion of the Alaskan center before 1968 and its subsequent contraction appears to be the primary reason for the 1967 break in the correlation between SATs at Barrow and St. Paul.

**Acknowledgments**

The author would like to thank two anonymous reviewers for their comments that helped to improve the manuscript.

**Conflicts of Interest**

*Climate* **2015**, *3* **491**